\documentclass[aps,prd,preprint,superscriptaddress,showpacs,showkeys]{revtex4}
\usepackage{amssymb,amsmath}
\usepackage{graphicx} 
\usepackage{dcolumn}  
\usepackage{epsfig}   
\usepackage{pstricks}
\usepackage{ifpdf}
\newcommand{\tr}{ {\mathrm{tr}\, }}
\newcommand{\Tr}{ {\mathrm{Tr}\, }}

\begin{document}

\preprint{INLN May 2012}

\title{Non-perturbative QCD amplitudes in quenched and eikonal approximations}


%
%

\author{H. M. Fried}
\affiliation{Physics Department, Brown University, Providence, RI 02912, USA}

\author{T. Grandou}
\affiliation{Universit\'{e} de Nice-Sophia Antipolis,\\ Institut Non Lin\'{e}aire de Nice, UMR 6618 CNRS 7335; 1361 routes des Lucioles, 06560 Valbonne, France}
\email[]{Thierry.Grandou@inln.cnrs.fr}

\author{Y.-M. Sheu}
\affiliation{Universit\'{e} de Nice-Sophia Antipolis,\\ Institut Non Lin\'{e}aire de Nice, UMR 6618 CNRS 7335; 1361 routes des Lucioles, 06560 Valbonne, France}


\date{\today}

\begin{abstract}
Even though approximated, strong coupling non-perturbative QCD amplitudes remain very difficult to obtain. In this article, in eikonal and quenched approximations, physical insights are presented that rely on the newly-discovered property of {\textit{effective locality}}.
\end{abstract}


\maketitle

\section{\label{SEC:1}Introduction}

As one derives the {\it{effective locality}} property of the fermionic QCD amplitudes~\cite{QCD1,QCD-II}, a bizarre factor of $\delta^{(2)}(b)$ comes about, in an unexpected place, where $b$ is the impact parameter of two scattering quarks.  Inspection shows that this curious factor is not related to any systematic approximation scheme, such as eikonal or quenching. Rather, it shows up as an unavoidable feature of any treatment, exact or approximate, and is likely to suggest the neglect of transverse fluctuations of quarks confined inside hadrons. In this respect, quite interesting perspectives offer themselves~\cite{QCD-II}.

Both quenched and eikonal approximations will be used throughout.  Both seem quite in line with the context of a highly energetic scattering process: The former because one may think that the vacuum back-reaction has not enough time to influence the process, and the latter because it is constructed in a Lorentz covariant way from the old Bloch-Nordsieck approximation~\cite{BN}, and is appropriate for high-energy scattering at momentum transfers substantially less than the scattering energies. For such processes, the Fourier momenta $k$ of the interacting $A_\mu(x)$- fields are typically of the same order-of-magnitude as the momentum transfer, and the conventional eikonal approximation is built around the idea that $k \ll E$ (for "hard scattering", there are eikonal variants of this "soft" scattering setting~\cite{HMF1}).

In this article, our aim is twofold. In the first place, it is made clear that this $\delta^{(2)}(b)$ can not be avoided, and a rigorous mathematical proof is proposed to support this claim. This is the matter of Section~\ref{SEC:2}. Then, in Section~\ref{SEC:3}, as a second step, in eikonal and quenched approximations, it is shown how the powerful {\textit{Random Matrix}} calculus can be used to obtain the generic, closed form expression of QCD fermionic amplitudes in the strong coupling regime. Eventually, in Section~\ref{SEC:4}, the forms so obtained will be used to revisit and improve upon a previous estimate of the quark/anti-quark binding potential and related aspects~\cite{QCD-II}.
\par
A discussion of the results is presented in Section~\ref{SEC:5}, and a short Appendix concludes this article.

\section{\label{SEC:2}The Constraint of delta-functions}

Using standard functional notations and manipulations~\cite{QCD1}, and a small re-arrangement that guarantees rigorous gauge invariance for all scattering quark amplitudes~\cite{QCD-II}, the QCD generating functional can be written as
\begin{eqnarray}\label{Eq:S2-1}
{Z}[j, \eta, \bar{\eta}] &=& \mathcal{N} \ e^{\frac{i}{2} \, \int{j \mathbf{D}_{c}^{(\zeta)}  j}} \int{\rm{d}[\chi]\   e^{\frac{i}{4} \int{\chi^{2}}}  \ e^{\mathfrak{D}_{A}} \  e^{\frac{i}{2} \, \int{\chi  \mathbf{F}} + \frac{i}{2} \int{A  {\mathbf{D}_{c}^{(\zeta)}}^{-1} A} } }\nonumber \\  && \times \left. e^{ i \int{\bar{\eta}  \mathbf{G}_{c}[A]  \eta}} \  e^{\mathbf{L}[A]} \right|_{A = \int{\mathbf{D}_{c}^{(\zeta)}  j}},
\end{eqnarray}

\noindent where $\mathcal{N}$ is a normalization constant, and the $j^\mu$, $\eta_\alpha$ and ${\bar{\eta}}_\beta$ stand for bosonic and fermionic sources, respectively. Note that the causal gluonic propagator $\mathbf{D}_{c}$ may differ by a factor of $2i$ from the more customarily used Feynman propagator.  The `Linkage Operator'
\begin{equation}
\mathfrak{D}_{A} = - \frac{i}{2} \int {\rm{d}}^4x\ {\rm{d}}^4y\ {\frac{\delta}{\delta A^{a}_{\mu}(x)}} \, \left. \mathbf{D}_{c}^{(\zeta)} \right|^{ab}_{\mu \nu}(x-y) \, \frac{\delta}{\delta A^{b}_{\nu}(y)}
\end{equation}

\noindent involves the covariant gluonic propagator with gauge parameter $\zeta$,
\begin{equation}
\left. \mathbf{D}_{c}^{(\zeta)} \right|^{ab}_{\mu \nu} = \delta^{ab} (-\partial^{2})^{-1} \left[ g_{\mu \nu} -(1- \zeta) {\partial_{\mu} \partial_{\nu}\over \partial^{2}} \right].
\end{equation}

\noindent The $\chi$-field appearing in $Z[j, \eta, \bar{\eta}]$ has long been used to {\it{linearize}} the exponential of the original $\mathbf{F}^2$-dependence that enters the original QCD Lagrangian density~\cite{Halpern1977,Reinhardt1993}:
\begin{equation}
e^{-\frac{i}{4} \, \int{ \mathbf{F}^{2}}} = {\mathcal{N}_\chi}\, \int{\mathrm{d}[\chi_{\mu\,\nu}^{a}]\, e^{{\frac{i}{4}}\, \int\chi^2 + {\frac{i}{2}}\, \int \chi^{\mu\, \nu}_{a}\, \mathbf{F}_{\mu\, \nu}^{a}} }.
\end{equation}

\noindent Proceeding in this way, the $A_\mu$-gauge field dependence are made gaussian instead of cubic and quartic, allowing for straight forward linkage operations.

In Eq.~(\ref{Eq:S2-1}), the original fermionic fields have been integrated out: This has given rise to the term involving the fermionic propagator $\mathbf{G}_c[A]$ in the gauge field background configuration $A$, as well as to the logarithm of the fermionic determinant $\mathbf{L}[A]$ (up to a constant  absorbed into the overall normalization  $\mathcal{N}$).  That is,
\begin{equation}
\mathbf{G}_{c}(x,y\, | A) = \langle x| [m + \gamma^{\mu}\, (\partial_{\mu} - i\, g\, A^{a}_{\mu}\, T_{a})]^{-1} | y\rangle\,
\end{equation}

\noindent and
\begin{equation}
\mathbf{L}[A] = \Tr{\ln[1 - i\, g\, (\gamma\, A\, T)\, \mathbf{S}_c]}, \quad \mathbf{S}_c = \mathbf{G}_c[0].
\end{equation}


In order to be able to display the property of effective locality, one must devise convenient enough representations for the functionals $\mathbf{G}_c[A]$ and $\mathbf{L}[A]$. Schwinger-Fradkin representations can be used; for example, a `mixed' (configuration and momentum space) expression of $\mathbf{G}_c[A]$ is given by

\begin{eqnarray}\label{Eq:S2-2}
{\langle p|\mathbf{G}_{c}[A] |y \rangle} &=&  e^{-i p \cdot y} \  i
\int_{0}^{\infty}{\mathrm{d}s \, e^{-is m^{2}}} \ e^{- \frac{1}{2} \Tr{\ln{\left( 2h \right)}} }\nonumber\\ \nonumber && \times \int{\mathrm{d}[u]} {\left\{ m - i \gamma \cdot \left[p - g A(y-u(s)) \right] \right\}} \ e^{{i\over 4} \int_{0}^{s}{\mathrm{d}s' \, [u'(s')]^{2} } } \ e^{i p \cdot u(s)}\nonumber\\  && \times  \left( e^{g \int_{0}^{s}{\mathrm{d}s' \sigma \cdot \mathbf{F}(y-u(s'))}} \ e^{-ig \int_{0}^{s}{\mathrm{d}s' \, u'(s') \cdot \mathbf{A}(y-u(s'))}} \right)_{+}
\end{eqnarray}

\noindent and similarly for $\mathbf{L}[A]$, \cite{QCD-II}. In the representation above, one has
\begin{eqnarray}
h(s_1,s_2)=s_1\Theta(s_2-s_1)+s_2\Theta(s_1-s_2)\ ,\ \ \ h^{-1}(s_1,s_2)=\frac{\partial}{\partial s_{1}} \frac{\partial}{\partial s_{2}} \delta(s_1-s_2),
\end{eqnarray}

\noindent and $\sigma_{\mu\nu}$, the customary commutator $\sigma_{\mu\nu}=\frac{1}{4}[\gamma_\mu,\gamma_\nu]$ controls the spin-related contributions.  This expression ends up with a subscript '$+$' to mean that time ordering with respect to Schwinger proper-time is in order, because of the $SU(N_c)$-Lie-algebra valuation of the gauge potentials $\mathbf{A}_\mu=A_\mu^aT^a$, and field-strength tensor $\mathbf{F}_{\mu\nu}=~F^a_{\mu\nu}T^a$.

The Fradkin field variables satisfy
\begin{equation}\label{Eq:S2-3}
\mathrm{u}_{\mu}(s) = \int_{0}^{s}{\mathrm{d}s' \, \mathrm{u'}_{\mu}(s')}, \quad \mathrm{u}_{\mu}(0) = 0,
\end{equation}

\noindent and a few words of explanation may be in order. From Fradkin's original representation, originally in terms of $\mathrm{v}_\mu(s')$ and here denoted by $\mathrm{u'}_{\mu}(s')$, this quantity has the interpretation of a particle's $4$-velocity or, depending on the mass-scale used, of a $4$-momentum.  But while undergoing fluctuations at every $s'$-value, that particle needs not be on its mass shell.  It is in the context of an eikonal approximation that $\mathrm{u'}_{\mu}(s')$ can be replaced by its mass shell $4$-momentum (that is, $\mathrm{u}_{\mu}(s)=sp_\mu$), because it is then assumed that the particle's $4$-momentum is much bigger than any momentum transfer ($|p-p'|/p \ll 1$), and that the mass shell hypothesis is accordingly a reliable one.  Note that this approximation should be relevant to the non-perturbative, semi-classical level of description considered here, whereas `harder' excitations should concern the next-to-leading order terms of a $\hbar$-expansion, that is, the description of the perturbative phase of QCD~\cite{deTeranmond2012}.

Eventually, it must be stressed that none of the features of effective locality is dependent upon this eikonal approximation; rather, the latter is used to provide a simple enough way to derive an effective quark-binding potential out of the relevant eikonal amplitude, taken as a function of the quark's {\textit{impact parameter}}, {\textit{i.e.}}, the inter-quark separation~\cite{QCD-II}.

The full derivation of the effective locality property will not be repeated here since its complete non-approximate proof can be found in Ref.~\cite{QCD-II}.  Rather, focus will be put on specific and important technical/physical aspects of effective locality.  On (\ref{Eq:S2-2}), however, and within the quenched approximation, it is immediate to realize that in order to be able to perform the linkage operations in an exact way, then, $\mathbf{G}_{c}[A] $ factors have to be pulled 'downstairs' by functional differentiations with respect to the fermionic sources $\eta$ and ${\bar{\eta}}$.  This is why the property of effective locality is not discernable from the generating functional itself, but after the sums over all gluon exchanges have been performed, on the full set of $2n$-point fermionic Green's functions.  Though equivalent, and thus not a crucial difference, a form of the effective action is under investigation, on which effective locality could be read off directly.  This point would be important when this formulation of QCD amplitudes generates a possible {\it{most dual}} QCD formulation, as was convincingly proposed in Ref.~\cite{Reinhardt1993} in the pure Yang-Mills case, or in relation to the recent results of Ref.~\cite{Ferrante2011}.
\par\medskip

In a first attempt to the effective locality property of QCD~\cite{QCD1}, and within the same approximate context as the one being studied in the present article, one finds in the effective action an exponential factor of
\begin{equation}\label{Eq:S2-5}
 - {i\over 2g} \int{\mathrm{d}^{4}x  \  Q^{a}_{\mu}(x)  \left. [f \cdot \chi(x)]^{-1}\right|^{\mu\nu}_{ab} \ Q^{b}_{\nu}(x)}
\end{equation}

\noindent representing, since all gluon interactions have been summed upon, the action of an effective local interaction term between two colored currents. The interacting currents are
\begin{equation}
 Q^{a}_{\mu}(x) =-\partial^{\nu} \chi^{a}_{\mu\nu}(x) + g [ R^{a}_{1 \mu}(x)+ R^{a}_{2 \mu}(x)],
\end{equation}

\noindent where $R^a_{i\mu}$, $i=1,2$, stand for their associated quark content. These expressions describe a $2$-body scattering process and is related to a $4$-point fermionic Green's function.

In eikonal approximation and in a non-perturbative strong coupling regime, $g\! \gg \!1$, the leading part of Eq.~(\ref{Eq:S2-5}) reads accordingly~\cite{QCD1,QCD-II}
\begin{eqnarray}\label{Eq:S2-6}
&& \frac{i}{2} g \int{\mathrm{d}^{4}w \, \int_{0}^{s}{\mathrm{d}s_{1} \, \int_{0}^{\bar{s}}{\mathrm{d}s_{2} \, \mathrm{u}'_{\mu}(s_{1}) \, \bar{\mathrm{u}}'_{\nu}(s_{2}) \,}}} \\ \nonumber && \quad \times \Omega^{a}(s_{1}) \, \bar{\Omega}^{b}(s_{2}) \,  \left. (f \cdot \chi)^{-1}(x) \right|^{\mu\nu}_{ab} \\ \nonumber && \times \, \delta^{(4)}(w_{} - y_{1} + \mathrm{u}(s_{1})) \, \delta^{(4)}(w_{} - y_{2} + \bar{\mathrm{u}}(s_{2})),
\end{eqnarray}

\noindent where the $\Omega^{a}(s_{1})$ and $ \bar{\Omega}^{b}(s_{2})$-variables are used so as to extract the $A_{\mu}^{a}$-field dependence out of the ordered exponential of Eq.~(\ref{Eq:S2-2})~\cite{QCD-II}.  The expression above is obtained in the standard eikonal and quenched approximations, which, for the former, also includes the neglect of spin-related contributions.   However, it is important to emphasize that the full non-approximate expression~\cite{QCD-II} would manifest exactly the same technical intricacy as the one under consideration, and this is why the point can be made using the simplified example of Eq.~(\ref{Eq:S2-6}).

The technically important aspect of (\ref{Eq:S2-6}) is the following: What is to be thought of
\begin{equation}\label{Eq:S2-7}
\delta^{(4)}(w_{1} - y_{1} + \mathrm{u}(s_{1})) \, \delta^{(4)}(y_{1} - y_{2} + \bar{\mathrm{u}}(s_{2}) - \mathrm{u}(s_{1}))?
\end{equation}

\noindent That is, how should one interpret such a factor as $\delta^{(4)}( C^{st}+ \bar{\mathrm{u}}(s_{2}) - \mathrm{u}(s_{1}))$ ?  At face value, for any given pair of values $(s_1, s_2)\in\  ]0, {s}]\times ]0, \bar{s}]$, and any pair of arbitrary functions $(\mathrm{u}, {\bar{\mathrm{u}}})$, each belonging to some infinite dimensional functional space, the probability of coincidence of $\mathrm{u}(s_1)$ with ${\bar{\mathrm{u}}}(s_2)+C^{st}$ is likely to be infinitesimally small, if not zero (the reason why the points $s_1=s_2=0$ are excluded will appear clear below, in Eq.~(\ref{Eq:mbym-2})).

Besides, in the $2$-body center of mass system ($CMS$), a somewhat heuristic manipulation of the $0$- and $3$- delta constraints would suggest that $C^{st}=0$, and write~\cite{QCD-II}
\begin{equation}\label{Eq:S2-8}
\delta(\bar{\mathrm{u}}_{0}(s_{2}) - \mathrm{u}_{0}(s_{1})) \, \delta( \bar{\mathrm{u}}_{3}(s_{2}) - \mathrm{u}_{3}(s_{1}))=\frac{\delta(s_{1}) \delta(s_{2})}{|\mathrm{u}'_{3}(0)| \, |\bar{\mathrm{u}}'_{0}(0)|} + \cdots ,
\end{equation}

\noindent where the dots are to be evaluated shortly. It is assumed that the Fradkin fields $\mathrm{u},\bar{\mathrm{u}}$ are ${C}^1 \left(]0,s], ]0, \bar{s}]\rightarrow {\mathbb{R}}^4\right)$. Then, out of $\delta^{(4)}( \bar{\mathrm{u}}(s_{2}) - \mathrm{u}(s_{1}))$, and in view of Eq.~(\ref{Eq:S2-8}), this leads to a remaining constraint of
\begin{equation}\label{Eq:S2-9}
\delta^{(2)}(\vec{y}_{1\perp} - \vec{y}_{2\perp}) \equiv {\delta^{(2)}(\vec{b})},
\end{equation}

\noindent where $\vec{b}$ is the impact parameter, or transverse distance between the two scattering quarks.  Given the very place where it comes about, as a multiplicative factor in an exponential, this ${\delta^{(2)}(\vec{b})}$ is awkward. In Ref.~\cite{QCD1}, it has been suggested that this factor is a remnant of the implicit existence of asymptotic quark states, an assumption which, beyond the stage of perturbation theory, cannot be maintained in QCD, neither theoretically nor experimentally~\cite{Lavelle1996}.

This point of view leads to a change of ${\delta^{(2)}(\vec{b})}$ into a modified, smeared and normalized impact parameter distribution of form~\cite{QCD-II}
\begin{equation}\label{Eq:S2-10}
{\delta^{(2)}(\vec{b})}\longrightarrow \varphi(b)=\frac{\mu^{2}}{\pi} \, \frac{1 + \xi/2}{\Gamma(\frac{1}{1+\xi/2})} \, e^{-(\mu b)^{2+\xi}} , \quad \xi\in{\mathbb{R}},\ \  |\xi| \ll 1,
\end{equation}

\noindent where $\mu$ is a typical mass term used to calibrate the transverse momenta distribution of quarks inside a given bound state.  In the transverse plane, bound quarks are endowed with transverse momenta taken as independent random variables, exponentially suppressed above a given mass parameter $\mu$.  That is, describing the non-perturbative regime of QCD, in order to avoid the nonsensical ${\delta^{(2)}(\vec{b})}$, it is necessary to introduce a mass scale where there wasn't any before, in agreement with the considerations of Ref.~\cite{BS2009}, for example. First fits, with respect to a model pion $Q{\bar{Q}}$ and a model nucleon $QQQ$~\cite{QCD-II}, indicate a value of $\mu$ close to the pion mass, and a value of $|\xi|={\sqrt{2}}/{{16}}$, on the order of $0.1$, small enough as it should if one is willing to preserve baryonic linear Regge behaviours through a(n almost) linear confining potential~\cite{deTeranmond2012}.

Eventually, it is worth remarking that taking $\xi$ negative-valued in Eq.~(\ref{Eq:S2-10}), then $\varphi(b)$ is (the {\textit{characteristic function}} of) a {\textit{L\'{e}vy-flights}} probability distribution (L\'{e}vy flight distributions are stable probability distributions).  They include and generalize Gaussian distributions for independent random variables, and comply with a generalized {\textit{central limit theorem}} of statistical physics~\cite{StableDistribution}). That is, starting from quark propagation as ordinarily conceived, \textit{\`a la} $\mathbf{G}_c(x,y|A)$, in the non-perturbative bound context, one could be lead to think of (transverse) quark propagation in terms of {\textit{L\'{e}vy-flights}} that are {\textit{markovian}} diffusive processes; a picture which, in view of color confinement, may look pretty sensible indeed.

Whatever its interpretation, though, the latter will be ruined if Eq.~(\ref{Eq:S2-8}) does not provide a reliable enough evaluation of Eq.~(\ref{Eq:S2-7}).  In order to deal with that issue, one may skip to the most achieved realization of a functional space, that is to the Wiener functional space~\cite{JohnsonLapidus}.  Note that even though that space may be not the only one to be conceived (and/or taken into account, according to Ref.~\cite{Ferrante2011}), there is no restriction at all in working out a proof within that space, because Fradkin vectorial fields are real valued.  Then, the following theorem can be proven:

\textit{{\bf{Theorem}}}: For all pairs $(s_1,s_2)\in\  ]0, {s}]\times ]0, \bar{s}]$,

\begin{equation}\label{Eq:measure}
m\otimes m\left(\bigl\lbrace(\mathrm{u}, \bar{\mathrm{u}})\in {C}^{0,{s}}_0\times {C}^{0,{\bar{s}}}_0\ | \ \mathrm{u}(s_1)=\bar{\mathrm{u}}(s_2)\bigr\rbrace\right) = 0,
\end{equation}

\begin{equation}\label{Eq:measure_normal}
m\otimes m\left(\bigl\lbrace(\mathrm{u}, \bar{\mathrm{u}})\in {C}^{0,{s}}_0\times {C}^{0,{\bar{s}}}_0\ | \ \mathrm{u}(0)=\bar{\mathrm{u}}(0)=0\bigr\rbrace\right)=1
\end{equation}

\noindent with the Wiener measure $m$ on ${C}^{0,{s}}_0$, whereas $m \otimes m$ is taken as the Wiener measure on the product of spaces ${C}^{0,{s}}_0\times {C}^{0,{\bar{s}}}_0$, endowed with the topology product. This is made possible thanks to the independence of the random variables $\mathrm{u}$ and $\bar{\mathrm{u}}$ that are here taken to represent either of the 2 possibilities $\mathrm{u}_{0,3}$ and $\bar{\mathrm{u}}_{0,3}$.

This theorem proves that the right hand side of Eq.~(\ref{Eq:S2-8}) is necessarily proportional to $\delta(s_1)\delta(s_2)$. Dimensional and symmetry arguments (the right hand side must be symmetric both in the exchange of indices $0\leftrightarrow 3$, and under the combined exchange $s_1 \leftrightarrow s_2$ and $\mathrm{u} \leftrightarrow {\bar{\mathrm{u}}}$, whereas the normalization is fixed by comparison to the eikonal approximation~\cite{QCD1}) can be used to complete the right hand side of Eq.~(\ref{Eq:S2-8}).  One finds
\begin{equation}\label{Eq:S2-11}
\delta(\bar{\mathrm{u}}_{0}(s_{2}) - \mathrm{u}_{0}(s_{1}))\delta( \bar{\mathrm{u}}_{3}(s_{2}) - \mathrm{u}_{3}(s_{1}))={1\over 2}\left({{ \delta(s_{1})\delta(s_{2})}\over |\mathrm{u}'_{3}(s_1)|  |\bar{\mathrm{u}}'_{0}(s_2)|}+{{ \delta(s_{1})\delta(s_{2})}\over |\mathrm{u}'_{0}(s_1)|  |{{\bar{\mathrm{u}}}}'_{3}(s_2)|}\right)
\end{equation}

\noindent and because of the $CMS$-relation $|\mathrm{u}'_{0}(s_1)||{{\bar{\mathrm{u}}}}'_{3}(s_2)|=|\mathrm{u}'_{3}(s_1)||\bar{\mathrm{u}}'_{0}(s_2)|$, which holds true at eikonal approximation, Eq.~(\ref{Eq:S2-11}) gets reduced to Eq.~(\ref{Eq:S2-8}) without the dots.

\par\medskip

The proof is as follows:

Let $A$ be the set $\bigl\lbrace(\mathrm{u}, \bar{\mathrm{u}})\in {C}^{0,{s}}_0\times {C}^{0,{\bar{s}}}_0\ | \ \mathrm{u}(s_1)=\bar{\mathrm{u}}(s_2)\bigr\rbrace$. One has $A=\bigcap_{n=1}^\infty A_n$, where
\begin{equation}
A_n=\bigl\lbrace(\mathrm{u}, \bar{\mathrm{u}})\in {C}^{0,{s}}_0\times {C}^{0,{\bar{s}}}_0\ | -{1\over n}\leq\ \mathrm{u}(s_1)-\bar{\mathrm{u}}(s_2)\leq +{1\over n}\bigr\rbrace.
\end{equation}

\noindent Because of the obvious inclusion, $\forall n$, $A_{n+1} \subset A_n$, one can write
\begin{equation}\label{Eq:mbym-1}
m\otimes m  (A)=\lim_{n \rightarrow \infty} m\otimes m (A_n).
\end{equation}

\noindent Now, $X_n\equiv m\!\otimes\! m (A_n)$ is given by
\begin{eqnarray}\label{Eq:mbym-2}
&& m\!\otimes\! m \left\{(\mathrm{u}, \bar{\mathrm{u}}) \subset {C}^{0,{s}}_0 \times {C}^{0,{\bar{s}}}_0\ |\  \mathrm{u}(s_1) -\frac{1}{n} \leq\ \bar{\mathrm{u}}(s_2)\leq \mathrm{u}(s_1)+\frac{1}{n} \, \right\} \nonumber \\ &=& \int_{-\infty}^{+\infty}{\frac{\mathrm{d}x} {\sqrt{2\pi s_1}} \, e^{-{x^2/ 2s_1}}} \, \int_{x-{1\over n}}^{x+{1\over n}}{\frac{\mathrm{d}y}{\sqrt{2\pi s_2}} \, e^{-{y^2/ 2s_2}}} \nonumber\\ &:=&  \int_{-\infty}^{+\infty}{\frac{\mathrm{d}x}{\sqrt{2\pi s_1}}\, e^{-\frac{x^2}{2s_1}} \, f_n(x)}.
\end{eqnarray}

\noindent Since one has the obvious inequality of
\begin{equation}
\left| \frac{1}{\sqrt{2\pi s_1}} \, e^{-\frac{x^2}{2s_1}} \, f_n(x) \right| \leq \frac{1}{\sqrt{2\pi s_1}} \, e^{-\frac{x^2}{2s_1}},
\end{equation}

\noindent one gets immediately
\begin{equation}
X_\infty=0
\end{equation}

\noindent as a mere consequence of the {\it{Dominated convergence theorem}}, allowing one to take the $n=\infty$-limit under the integral, whereas the second equation, Eq.~(\ref{Eq:measure_normal}), follows from normalization.

The unavoidable presence of this surprising factor of ${\delta^{(2)}(\vec{b})}$ is thus demonstrated. Then, in order to get a non-trivial (almost linear) inter-quark binding potential at moderate distance, no much larger than $\mu r=1$, (to be probed, larger distance requires the inclusion of quark loops~\cite{QCD5}, whereas Eq.~(\ref{Eq:S2-12}) below is obtained with the full quenched approximation),
\begin{equation}\label{Eq:S2-12}
V(r)\simeq - \xi \mu(\mu r)^{1+\xi},\ \ \ \ \ {\rm{at}}\ \  \mu r \geq 1,
\end{equation}

\noindent a smearing of the initial ${\delta^{(2)}(\vec{b})}$ in the form prescribed by Eq.~(\ref{Eq:S2-10}) is mandatory~\cite{QCD-II}, with that small deformation parameter $\xi$.  Fascinating relations ensue thereof, between L\'evy flight modes of propagation for confined quarks (at $\xi<0$), chiral symmetry breaking, confinement and an unexpected non-commutative geometrical aspect of the scattering transverse planes ({\textit{de Moyal}} planes~\cite{Jackiw}). These relations shall be discussed elsewhere.

\section{\label{SEC:3}Fermionic QCD amplitudes in eikonal and quenched approximations}

As a result of effective locality, one finds a 2-body scattering amplitude proportional to
\begin{eqnarray}\label{Eq:1}
&& \int{\mathrm{d}^n \alpha_1} \int{\mathrm{d}^n \alpha_{2} } \int{\mathrm{d}^n \Omega_1} \int{\mathrm{d}^n \Omega_{2} \, e^{-i\alpha_1\cdot\Omega_I}\, e^{-i\alpha_{2}\cdot\Omega_{2}}\, e^{i\alpha_1\cdot T}\ e^{i\alpha_{2}\cdot T} } \nonumber \\ && {\cal{N}}\prod^{1\leq a\leq n}_{ 0\leq\mu<\nu\leq 3}\int{\mathrm{d}[\chi^a_{\mu\nu}(w)] \, \det[gf\cdot\chi(w)]^{-{1\over 2}}\ e^{{i\over 4}\chi^2(w)+ig\varphi(b)\ \Omega_1^a[f \cdot \chi(w)]^{-1}\vert^{ab}_{30}\ \Omega^b_{2}} },
\end{eqnarray}

\noindent where $n$ is the shorthand for $N_c^2-1$, and the dot, ` $\cdot$', denotes an Euclidean scalar product in a relevant representation space; for example, the product $[f\cdot\chi]^{bc}=\sum_af^{abc}\chi^a$. The $T^a$s are the $n$ hermitian traceless generators of the $SU(N_c)$-Lie-algebra, taken in the fundamental $ N_c$-dimensional representation of $SU(N_c)$. In the adjoint representation to be used shortly, the $T^a$s are given by the structure constants of the $SU(N_c)$-Lie-algebra: $(T^a)_{bc}=-if_{abc}=-if^{abc}$. The form of Eq.~(\ref{Eq:1}) is non-trivial and is justified in Ref.~\cite{QCD1}. The full 2-body amplitude requires several prefactors that are not written in Eq.~(\ref{Eq:1}) since they have no relevance to the consideration to follow.

In the second line of Eq.~(\ref{Eq:1}), the last exponential argument is nothing else than Eq.~(\ref{Eq:S2-6}) evaluated with the help of identity Eq.~(\ref{Eq:S2-11}).  In eikonal approximation, one gets in effect
\begin{eqnarray}\label{Eq:2}
\exp{\left\{ {i\over 2}g\varphi(b){u_\mu'(0){\bar{u}}'_\nu(0)} \, \frac{1}{2}\ \left[ \frac{1}{|u'_0(0)||{\bar{u}}'_3(0)|} + \frac{1}{|u'_3(0)||{\bar{u}}'_0(0)|} \right] \, [f \cdot \chi(w)]^{-1} \vert_{ab}^{\mu\nu}\ \Omega_1^a(0)\Omega^b_{2}(0)\right\}},
\end{eqnarray}

\noindent where $u'(0) = p_{1} $ and ${\bar{u}}'(0) = p_{2} $, and where the original $\delta^{(2)}(b)$ has been turned into the distribution $\varphi(b)$ of Eq.~(\ref{Eq:S2-10}).  In the $CMS$ of the two scattering quarks of momenta $p_1$ and $p_2$, Eq.~(\ref{Eq:2}) reduces to the last argument in the exponential at the end of the expression in Eq.~(\ref{Eq:1}).

It is worth stressing how the peculiar form of Eq.~(\ref{Eq:1}) originates from the property of effective locality. Because of Eq.~(\ref{Eq:S2-7}), one has $w=(0,\vec{y}_{\perp},0)_{CMS}$ with $\vec{y}_{\perp}=\vec{y}_{1 \perp}=-\vec{y}_{2 \perp}=\frac{1}{2}{\vec{b}}$ as the only spacetime point where the interaction takes place, and the contributions attached to all the other points are eliminated. The net result is that there is no spacetime integration in the argument of the exponential at the end of Eq.~(\ref{Eq:1}); and accordingly, no spacetime integration either for the first term.

This can be understood in a straight forward way by recalling the standard procedure of a generating functional construction~\cite{TG2013}, where the spacetime manifold is broken up into an infinite series of infinitesimal cells.  Since the interaction, of the contact type, takes place at a unique point, $w$, the contributions of all of the other infinitesimal cells, centered at different points, just contribute a product of infinite multiplicative factors of $1$, thanks to the normalization of each cell.

%
%
%
%



%
%

\par\smallskip
The final result for the amplitude will therefore depend on a length scale that is not required by the interaction term itself, since it is dimensionless, but rather by the action term $\exp{[\frac{i}{4} \int{ {\rm{d}}^{4}z \, \chi^{2}(z)} ]}$, which, because of effective locality, is now to be necessarily understood as $\exp{[\frac{i}{4} \delta^4 \chi^2(w)]}$. A non-trivial result will depend upon forming a connection between the infinitesimal mathematical $\delta$ and a physically meaningful $\delta_{\mathrm{ph}}$ related to the scattering process under consideration~\cite{Schwinger}.

For such a 2-body process, the standard laws of Quantum Mechanics lead to a cell of time extension of order $1/E$, while that corresponding to a high-energy $CMS$ longitudinal coordinate should be $1/p_{\mathrm{L}} \simeq 1/E$; eventually, transverse coordinates should extend over a distance of order $1/\mu$ each, \textit{c.f.} Eq.~(\ref{Eq:S2-10}), and the overall interaction cell volume is therefore on the order of $\delta_{\mathrm{ph}}^{4} = (1/E\mu)^{2}$.

Reciprocally, the dimensionless interaction term is now re-written as
\begin{equation}
ig\varphi(b)\ \Omega_1[f\cdot\chi]^{-1}(w)\ \Omega_2\longrightarrow ig\ [\varphi(b) \delta_{\mathrm{ph}}^{2}] \ \Omega_1 \ [f \cdot \bar{\chi}(w)]^{-1}\ \Omega_2,
\end{equation}

\noindent where the field variables ${\bar{\chi}}=\delta_{\mathrm{ph}}^{2}\chi$ are dimensionless, as is the new combination
\begin{equation}\label{newphi}
\varphi(b)\,\delta_{\mathrm{ph}}^{2}=C^{st}\,({\mu/{\sqrt{\hat{s}}}})\,e^{-(\mu b)^{2+\xi}},
\end{equation}

\noindent where $\hat{s}=(p_1+p_2)^2$. Hereafter, the bar over $\chi$-fields will be omitted; and instead of Eq.~(\ref{Eq:S2-10}), the right hand side of Eq.~(\ref{newphi}) will be understood as $\varphi(b)$, which is now dimensionless.


%
%
%

As it stands, however, the expression Eq.~(\ref{Eq:2}) confronts us with a puzzling issue, that is inverting the quantity $[f\cdot\chi]$.  Taken as some relevant matrix, in effect, it is immediate to realize that zero belongs to the spectrum of $[f\cdot\chi]$, so that there seems to be no way out of this dead end.  However, two most favorable circumstances allow one not only to cope with that issue, but also to reach the level of actual, sensible calculations.

The first one of these two circumstances is as follows.  In order to:
\begin{enumerate}
  \item Invert, at least formally (for the time being) the product $[f\cdot\chi]$,
  \item and take into account the contributions of all of the $6n$- $\chi^a_{\mu\nu}$-fields entering $[f\cdot\chi]$ at the spacetime dimension of $D=4$ (contrast to Ref.~\cite{QCD1},  where, for simplicity, only $n$ components $\chi^a_{03}$ were considered),
  \item as well as the full symmetry of the product $[f \cdot \chi(w)]$ under the combined exchanges, $a \leftrightarrow b$ and $\mu \leftrightarrow \nu$~\cite{Reinhardt1993},
\end{enumerate}

\noindent it is appropriate to think of $(f\cdot\chi)$ as a tensorial product $\sum_{a=1}^n{\chi^a}_{\mu  \nu}\otimes T^a$ of operators acting on the product space $\{\mu,\nu\}\otimes \{a,b\}$. The first space, $\{\mu,\nu\}$, is the {\it{vectorial}} Minkowski space (as opposed to the {\it{affine}} Minkowski space, the space of event-points, to which the former space is tangent), and the second, $\{a,b\}$, a $SU(N_c)$-adjoint representation space with dimension $n= N_{c}^{2}-1$.  Their product is a $Dn\equiv N$-dimensional vectorial space that can be endowed with the product basis $\{\varphi_1, \varphi_2, \dots, \varphi_{N}\}$ constructed out of the bases of $\{\mu,\nu\}$ and $\{a,b\}$ spaces, on which act the 2-form  ${\chi^a}_{\mu\nu}\,\epsilon^\mu\wedge\epsilon^\nu$ and the endomorphism $T^a$, respectively.

In this product space, at real-valued Halpern-fields, and with $f^{abc}=i T^a_{bc}$, the operator $(\sum_{a=1}^n{\chi^a}\otimes T^a)$ is represented by a $N\times N$ matrix equal to $i$ times a real symmetric and traceless matrix $M$,
\begin{equation}\label{Eq:4}
 \sum_{a=1}^n{{\chi^a}\otimes T^a} = iM\, ,\  \quad M_{ij}= M_{ji}\in \mathbf{R}\, ,\ \ 1\leq i,j\leq N\, ,\quad \tr M=0.
\end{equation}

\noindent The operator $\chi \otimes T$ acts upon the $N$-component vectors $u'(0)\otimes \Omega_1(0)\equiv V_1$, ${\bar{u}}'(0)\otimes \Omega_2(0)\equiv V_2$ of the product space, and in the basis $\{\varphi_1, \varphi_2, \dots, \varphi_{N}\}$, these vectors have components as
\begin{equation}\label{Eq:5}
V_1={}^t(u'_0 \Omega^1_1, u'_0 \Omega^2_1, \dots, u'_3 \Omega^n_1), \quad V_2={}^t({\bar{u}}'_0 \Omega^1_2, {\bar{u}}'_0 \Omega^2_2, \dots, {\bar{u}}'_3 \Omega^n_2),
\end{equation}

\noindent where, for short, the dependence on $s_i=0$, $i=1,2$, are hereafter omitted, and where the superscript '$t$' stands for `transpose'.  Note that unlike the $\Omega_i$s, the $V_i$s are not dimensionless objects since, as quoted before, $u'(0)=p_1$ and ${\bar{u}}'(0)=p_2$. Note also that, out of respective scalar products of the 2 spaces, a scalar product on the product space $\{\mu,\nu\}\otimes \{a,b\}$ can be defined in a canonical way, and inherits the pseudo-Euclidean character of the Minkowskian $\{\mu,\nu\}$ space (this is exploited in the calculations of Section~\ref{SEC:4}).

Then, there exists an orthogonal matrix ${\cal{O}}$ that takes $M$ to its diagonal form
\begin{equation}\label{O}
{}^t{\cal{O}} M\,{\cal{O}}=\mathrm{diag} \, (\dots ,\xi_i,\dots), \quad i=1,2,\dots, N, \quad \xi_i \in \mathrm{Sp}(M)\subset \mathbf{R},
\end{equation}

\noindent where ${}^t{\cal{O}}$ is the transpose of ${\cal{O}}$, and likewise, $M^{-1}$ to its diagonal form also
\begin{equation}\label{Eq:6}
{}^t{\cal{O}}M^{-1}{\cal{O}}=\mathrm{diag} \, (\dots ,\frac{1}{\xi_i},\dots), \quad i=1,2,\dots, N.
\end{equation}

\noindent The spectrum of $M$ may contain the eigenvalue $\xi_i=0$, so that, for the time being, Eq.~(\ref{Eq:6}) can only be given a formal meaning. Re-writing Eq.~(\ref{Eq:2}) as
\begin{equation}\label{Eq:7}
\exp\left( -i\frac{g\varphi(b)}{2Ep}\ V_1^i\ [M^{-1}]_{ij}\ V_2^j\right),
\end{equation}

\noindent and redefining the $V_i$-fields,
\begin{equation}\label{Eq:8}
V_i\rightarrow V_i'={\cal{O}}V_i,
\end{equation}

\noindent one gets for Eq.~(\ref{Eq:7}) the expression
\begin{equation}\label{Eq:9}
\exp\left( -i\frac{{g\varphi(b){}}}{2Ep}\sum_{i=1}^{N}{ \frac{{V'_1}^i {V'_2}^i}{\xi_i} }\right).
\end{equation}

\noindent At eikonal approximation, fixed $u',{\bar{u}}'=p_1,p_2$, one has ${\rm{d}}^n\Omega_i= m^{-N}{\rm{d}}^{N}(p_i\otimes\Omega_i)$; and likewise, a trivial Jacobian for the transformation Eq.~(\ref{Eq:8}).  By using the reformulation Eq.~(\ref{Eq:4}), the integration with measure
\begin{equation}
\prod^{1\leq a\leq n}_{ 0\leq\mu<\nu\leq 3} \int {\rm{d}}[ \chi^a_{\mu\nu}(w)],
\end{equation}

\noindent which appears in Eq.~(\ref{Eq:1}), gets translated into an integration over the algebra of real symmetric $N\!\times\! N$ traceless matrices with measure~\cite{Mehta1967}
\begin{eqnarray}\label{Eq:10}
& & {\rm{d}}(\sum_{a=1}^n{\chi^a}_{\mu \nu}\otimes T^a) \nonumber \\ &=& {\rm{d}}M= {\rm{d}}M_{11}\,{\rm{d}}M_{12} \cdots {\rm{d}}M_{NN} \nonumber \\ &=&\nonumber \left|\frac{ \partial(M_{11}, \cdots, M_{N\!N})}{\partial(\xi_1, \cdots, \xi_N, p_1, \cdots, p_{N(N-1)/2})}\right| \, {\rm{d}}\xi_1 \cdots {\rm{d}}\xi_N \, {\rm{d}}p_1 \cdots {\rm{d}}p_{N(N-1)/2} \\  &=& \prod_{i=1}^{N}\ {\rm{d}}\xi_i  \prod_{1\leq i<j\leq N} |\xi_i-\xi_j|\   {\rm{d}}p_1\  ..\ {\rm{d}}p_{N(N-1)/2}\, f(p).
\end{eqnarray}

\noindent The $N$ eigenvalues of $M$ are real, and $f(p)$ is a(n unspecified) function of the $N(N-1)/2$ extra parameters which complete the parametrization of a given real symmetric matrix $M$, and fully determine the orthogonal matrix ${\cal{O}}\in SO_N(\mathbf{R})$.  Mathematically, this passage from an infinite dimensional functional space of integration, to a finite dimensional matrix space is non-trivial; it is a direct consequence of the effective locality and can be proven by relying on the {\it{measure image theorem}}~\cite{JohnsonLapidus,CandelpergherGrandou}.  The dependences on the extra parameters $p_l, l=1,\dots, N(N-1)/2$ of the orthogonal matrices ${\cal{O}}(p)$ will play a role in the calculations of the next Section~\ref{SEC:4}.

The second line of Eq.~(\ref{Eq:1}) can accordingly be re-written as
\begin{eqnarray}\label{Eq:11}
g^{-\frac{N}{2}}\ {\cal{N}}\int^{+\infty}_{-\infty}[\prod_{i=1}^{N}\ {{\rm{d}}\xi_i}]\,  [\prod_{1\leq i<j\leq N} |\xi_i-\xi_j|\ ]
\ e^{-{i\over 8N_c}\, \xi_i^2}\ \ {\frac{1}{\sqrt{\xi_i}}}\,{e^{-i{g\frac{\varphi(b)}{2Ep}}\ { \frac{{V'_1}^i{V'_2}^i}{\xi_i}}}},
\end{eqnarray}


\noindent where, for the first exponential term, the relation
\begin{equation}
\sum_{a=1}^{N_c^2-1}\chi^a_{\mu\nu}{\chi^a}^{\mu\nu}=(2N_c)^{-1}\,\tr M^2
\end{equation}

\noindent has been used.  Keeping in line with Ref.~\cite{Mehta1967}, the normalization constant ${\cal{N}}$ may be so defined as to normalize Eq.~(\ref{Eq:11}) to unity when the whole set of functions
\begin{eqnarray}\label{Eq:11'}
f^{(4)}_i(\xi_i)=\frac{1}{\sqrt{g}}  {1\over {\sqrt{\xi_i}}}\, e^{-\frac{B^{(i)}_4}{\xi_i}}
\end{eqnarray}

\noindent is replaced by the set of constant unit functions, $f^{(4)}_i(\xi_i)= 1$ (by the way, this turns out to be the normalization used in Eq.~(40) of Ref.~\cite{QCD1}, and the superscript $(4)$ refers to a $4$-point function).
\par\medskip
In order to allow for a concise and closed form expression, one can take advantage of the method used in Ref.~\cite{Mehta1967}; a simple re-scaling of the eigenvalues is necessary, that is, $\xi'_i={\sqrt{{1/ 4N_c}}}\,\xi_i$, taking Eq.~(\ref{Eq:11}) to the form relevant to the {\textit{Gaussian Orthogonal Ensemble}}~\cite{Mehta1967},
\begin{eqnarray}\label{Eq:12}
g^{-N\over 2}({4N_c})^{{N^2\over 2}-{N\over 4}}\ {\cal{N}}\int^{\infty}_{-\infty}\prod_{i=1}^{N}\ {d\xi_i\over {\sqrt{\xi_i}}}  \prod_{ i<j} |\xi_i-\xi_j| \ e^{-\frac{i}{2}\,\xi_i^2}\ e^{-i{g\varphi(b)\over 2Ep}{\sqrt{{1\over 4N_c}}}\,{ {V'_1}^i{V'_2}^i\over \xi_i}},
\end{eqnarray}

\noindent in which appears a {\it{Vandermonde determinant}} of
\begin{eqnarray}\label{Eq:13}
\prod_{1\leq i<j\leq N} |\xi_i-\xi_j| = {\cal{P}}(\xi_1,\dots,\xi_{N})
\end{eqnarray}

\noindent with ${\cal{P}}(\xi_1,\dots,\xi_{N})$, the absolute value of a sum of $2^{N(N-1)/2}$ monomials in $\xi_i$ of overall degree $N(N-1)/2$.  The expression
\begin{eqnarray}\label{Eq:15}
P_{N_1}(\xi_1,\dots,\xi_{N})=C_{N_1}\ e^{-\frac{i}{2}\sum \xi_i^2}\,{\cal{P}}(\xi_1,\dots,\xi_{N})
\end{eqnarray}




\noindent is normalized so that its integration over the full spectrum of eigenvalues $\xi_i$, $1\leq i\leq N,$ yields unity, and one gets~\cite{Mehta1967}
\begin{eqnarray}\label{Eq:18}
C_{N_1}^{-1}=2^{\frac{3N}{2}}\,\prod_{j=1}^{N} \Gamma(1+\frac{j}{2}),
\end{eqnarray}

\noindent allowing one to identify the normalization constant of Eq.~(\ref{Eq:12}) as
\begin{equation}\label {N}
{\cal{N}}=C_{N_1}\, ({4N_c})^{-{N^2\over 2}+{N\over 4}}.
\end{equation}

\noindent Things may be expressed in terms of harmonic oscillator wave functions,
\begin{eqnarray}\label{Eq:14}
\varphi_k(x)={1\over {\sqrt{2^kk!{\sqrt{\pi}}}}}\ e^{-{1\over 2}x^2}\ H_k(x),
\end{eqnarray}

\noindent where $H_k(x)$ is the Hermite polynomial of order $k$.  Their normalization reads as
\begin{equation}\label{Eq:normal}
\int_{-\infty}^{+\infty}{\rm{d}} x\  \varphi_k(x) \varphi_j(x)=\delta_{kj} .
\end{equation}

\noindent However, contrary to the standard expression of Ref.~\cite{Mehta1967}, Eq.~(\ref{Eq:15}) displays a factor of $i$ in its exponential term, which, in Ref.~(\ref{Eq:normal}), translates into a ${\sqrt{i}}\xi_j$-dependence of the harmonic oscillator wave functions, rather than just $\xi_i$. Relying on Cauchy's Theorem, though, it is proven in the Appendix that the integration formulae of Eqs.~(\ref{Eq:15}) and (\ref{Eq:normal}) defined without ${\sqrt{i}}$, can be analytically continued from real valued $\xi_j$ to the complex range of $\{z| \, z=\xi_je^{i\theta},\ 0\leq\theta\leq \pi/4\}$. In particular, using this analytical continuation it is possible to check that the harmonic oscillator wave function normalization just undergoes a simple re-scaling of
\begin{eqnarray}\label{Eq:19}
\int_{-\infty}^{+\infty}{\rm{d}} x\  \varphi_k(x) \varphi_j(x)=\delta_{kj}\longmapsto \int_{-\infty}^{+\infty}{\rm{d}} x\  \varphi_k(x{\sqrt{i}}) \varphi_j(x{\sqrt{i}})=\frac{1}{\sqrt{i}} \, \delta_{kj}.
\end{eqnarray}

\noindent Then, proceeding along the standard steps of Ref.~\cite{Mehta1967}, one can check that Eq.~(\ref{Eq:12}) can be simply written as
\begin{equation}\label{Eq:20}
{\sqrt{i}}\,C_{N_1}\,\prod_{i=1}^{N}\int^{+\infty}_{-\infty} {{\rm{d}}}\xi_i\  \ e^{-\frac{i}{2}\,\xi_i^2}\prod_{1\leq i<j\leq N} |\xi_i-\xi_j|\, \ f_i^{(4)}(\xi_i)\, .
\end{equation}

The absolute values in Eqs.~(\ref{Eq:13}) and (\ref{Eq:20}) are bothering. At a formal level at least, Eq.~(\ref{Eq:20}) can be expressed in terms of normalized harmonic oscillator wave functions under the form
\begin{eqnarray}\label{Eq:22}
& & {{\sqrt{i}}}   \, C_{N_1} N!(\prod_{j=0}^{N-1} {\sqrt{2^jj!{\sqrt{\pi}}}})\prod_{i=1}^{N}\int^{+\infty}_{-\infty}\!\! {{\rm{d}}}\xi_i\prod_{ i<j}^N \varepsilon(\xi_i-\xi_j)\, f_i^{(4)}(\xi_i)\det [\,\varphi_{k-1}(\xi_l)\,], \\ \nonumber & & \quad \quad \quad k, l =1,..,N,
\end{eqnarray}

\noindent where $\varepsilon(x)$ is the distribution {\textit{sign of x}}, and $f_i^{(4)}(\xi_i)$ are those of Eq.~(\ref{Eq:11'}) with
\begin{eqnarray}\label{Eq:25}
B^{(i)}_4=i{g{\varphi(b) }\over {\sqrt{{16N_c}}}}\,{{V'}_1^i{V'}_2^i\over Ep}\,,
\end{eqnarray}

\noindent where the rescaling from Eqs.~(\ref{Eq:11}) to (\ref{Eq:12}) has been taken into account.

It is remarkable that exactly the same equations as Eqs.(\ref{Eq:11'}), (\ref{Eq:20}) and (\ref{Eq:22}) express the closed form of any $2m$-point fermionic amplitudes (not yet integrated over the extra auxiliary variables $\alpha_i$ and $\Omega_i$ of Eq.~(\ref{Eq:1}); see next Section~\ref{SEC:4}).  In the general case of a fermionic $2m$-point function, in effect, all of the previous steps can be followed through, and the only change turns out to be the redefinition of the constant $B^{(i)}_4$.  Within obvious notations ($|p_k|=|p_l|=p_{kl}$, and $E_k=E_l=E_{kl}$, in each of the $2$ by $2$-$`kl'\!-\!CMSs$),
\begin{eqnarray}\label{Eq:25'}
B^{(i)}_4\longmapsto B^{(i)}_{2m}=i{g\over {\sqrt{{16N_c}}}} \sum_{1\leq k<l\leq m}{\varphi(b_{kl}) \over E_{kl}\,p_{kl}}\, {V'}^i_k{V'}^i_l ,
\end{eqnarray}

\noindent where, again, the restriction $k<l$ in the summation discards the self-energy effect that will be dealt with in another article. In Ref.~\cite{QCD5}, though, a glance at this issue is obtained: Amazingly, the same ultraviolet renormalization structure as in the usual QCD perturbative regime comes about, and it is most interesting to note that this point is in line with the explanation proposed in Ref.~\cite{deTeranmond2012}, based on an analysis of the AdS/QCD correspondence.

To summarize, at quenched and eikonal approximations, and up to renormalization, a $2m$-point fermionic amplitude can concisely be written under a form proportional to
\begin{eqnarray}\label{Eq:26}
&& \int{\mathrm{d}^n \alpha_1} \int{\mathrm{d}^n \alpha_{2} } \int{\mathrm{d}^n \Omega_1} \int{\mathrm{d}^n \Omega_{2} \, e^{-i\alpha_1\cdot\Omega_I}\, e^{-i\alpha_{2}\cdot\Omega_{2}}\, e^{i\alpha_1\cdot T}\ e^{i\alpha_{2}\cdot T} } \nonumber \\ && {\sqrt{i}}\,C_{N_1}\,\prod_{i=1}^{N}\int^{+\infty}_{-\infty} {{\rm{d}}}\xi_i\  \ e^{-\frac{i}{2}\,\xi_i^2}\prod_{1\leq i<j\leq N} |\xi_i-\xi_j|\, \ f_i^{(2m)}(\xi_i).
\end{eqnarray}

\noindent Unfortunately, the {\textit{method of integration over alternate variables}} of Ref.~\cite{Mehta1967}, aimed at circumventing the absolute values of Eqs.~(\ref{Eq:20}) and (\ref{Eq:26}), is able to yield elegant closed form expressions such as (\ref{Eq:22}), but is of no real help to calculational purposes.





%
%
%

\section{\label{SEC:4}$G^{mn}_{pq}$-Meijer's special functions}

The general and closed form expressions given above may be most appropriate to numerical simulations. In this section though, for the sake of illustration and a further speculation on non-perturbative QCD, it is enough to get back to an expression close to Eq.~(\ref{Eq:20}),
\begin{equation}\label{56}
{\sqrt{i}}\,C_{N_1}\,\prod_{i=1}^{N}\int^{+\infty}_{-\infty} {{\rm{d}}}\xi_i\  \ e^{-\frac{i}{2}\,\xi_i^2}\prod_{1\leq i<j\leq N} (\xi_i-\xi_j)\, \ f_i^{(4)}(\xi_i)\, ,
\end{equation}

\noindent where, anticipating somewhat on a result that does not depend on them, the prescriptions of absolute values have been given up.  Likewise, in this section, calculations are restricted to the $4$-point fermionic function, that is $m=2$, and the superscript $(4)$ will hereafter be omitted for short. The Vandermonde determinant being a polynomial in eigenvalues, one has to cope with generic integrations of type
\begin{eqnarray}\label{Eq:27}
\int^{+\infty}_{-\infty}{\frac{\mathrm{d}\xi_{i}}{\sqrt{\xi_i}} \, \xi_i^{q_i} \, e^{-\frac{i}{2} \xi_i^2- \frac{B^{(i)}}{\xi_i}}}={\sqrt{2}}^{q_i+\frac{1}{2}}\int^{+\infty}_{-\infty}{{\mathrm{d}\xi_{i}} \ \xi_i^{q_i-\frac{1}{2}} \, e^{-i \xi_i^2- \frac{B^{(i)}/{\sqrt{2}}}{\xi_i}}},
\end{eqnarray}

\noindent where $q_i \in \mathbb{I\!N}$. Defining

\begin{eqnarray}\label{Eq:28}
I_p(b)=\int_0^\infty\ {{\rm{d}}x}\ x^p\ e^{-x^2-{b\over x}}
\end{eqnarray}

\noindent for $p>0$ and $b>0$, one has that Eq.~(\ref{Eq:28}) evaluates to~\cite{Apelblat1983}
\begin{eqnarray}\label{Eq:29}
I_p(b)={\frac{1}{2 \sqrt{\pi}}}\, G^{30}_{03}\left(\frac{b^2}{4} \biggr| {p+1\over 2}, \frac{1}{2},0\right),
\end{eqnarray}

\noindent where $G^{30}_{03}(x|b_1,b_2,b_3)$ is a {\textit{Meijer's special function}}, and the second of the two most favorable circumstances alluded to, after Eq.(\ref {newphi}), is the following: Whereas Eq.~(\ref{Eq:29}) holds true at values of $p$ and $b$ such as specified above, it turns out that $G^{30}_{03}(x|b_1,b_2,b_3)$ is analytic in $x$, and that the parameters appearing after $x$ can also be continued to complex values~\cite{Erdelyi1953}.

On the basis of these analyticity properties, it is therefore possible to give the generic integrations of Eq.~(\ref{Eq:27}) the meaning of
\begin{eqnarray}\label{Eq:30}
\int^{+\infty}_{-\infty}\! {{\rm{d}}\xi_i\over {\sqrt{\xi_i}}}\ \xi_i^{q_i}\ e^{-{i\over 2}\xi_i^2-\frac{B^{(i)}/{\sqrt{2}}}{\xi_i}}={\sqrt{{-4i}}}^{q_i+{1\over 2}}{[1-i(-1)^{q_i}]\over 2{\sqrt{\pi}}}G^{30}_{03}\left({i(B^{(i)})^2\over 8} \biggr| {2q_i+1\over 4}, {1\over 2},0\right),
\end{eqnarray}

\noindent where the equality, demonstrated in the Appendix, is a consequence of {\textit{Cauchy's Theorem}}, and where the constants $B^{(i)}$, $i=1, \dots, N$, are defined in Eq.~(\ref{Eq:25}).

Now, each of the $B^{(i)}$ is on the order of $g\varphi(b)$, and the small $g\varphi(b)$ limit can therefore be investigated on the basis of the small $z$ expansion, $|z| \ll 1$,
\begin{eqnarray}\label{Eq:31}
G^{30}_{03}\left(z| b_1,b_2, b_3\right)=\sum_{h=1}^3{{\prod'}_{j=1}^3\Gamma(b_j-b_h)\over  \Gamma(1+b_h-b_3)} \ z^{b_h}\times {}_0F_2[\ ..\ ; -z],
\end{eqnarray}

\noindent where the prime on top of the product indicates that the pole value $b_j=b_h$ is omitted, and likewise the asterisk in the ${}_0F_2$ generalized hypergeometric series, indicates also that the value $b_j=b_h$ is omitted:
\begin{eqnarray}\label{Eq:32}
{}_0F_2[1+b_h-b_1,\dots,*,\dots1+b_h-b_3;-z].
\end{eqnarray}

\noindent For all $b_h$, one has, at $|z| \ll 1$,
\begin{eqnarray}\label{Eq:33}
{}_0F_2[1+b_h-b_1,\dots,*,\dots1+b_h-b_3;z]=1+{\cal{O}}(z).
\end{eqnarray}
\noindent
For the situation considered in Eq.~(\ref{Eq:29}), the parameter sets $(b_1, b_2, b_3)$ always read as $\left(\frac{2q_i+1}{4}, \frac{1}{2}, 0\right)$, so that in the small $z$-limit, the Meijer's special function (\ref{Eq:29}) yields, for example,
\begin{eqnarray}
&& G^{30}_{03}\left(i{(B^{(i)})^2\over 8}| {2q_i+1\over 4}, {1\over 2},0\right) \nonumber= \\ \nonumber &&  \left\{ \Gamma({1\over 2})\Gamma({2q_i+1\over 4})+{\Gamma(-{1/ 2})\Gamma({(2q_i-1)/ 4})\over \Gamma(3/2)}[i{(B^{(i)})^2\over 8}]^{1\over 2} \right. \\  && \left. + {\Gamma(-{(2q_i+1)/ 4})\Gamma({-(2q_i-3)/ 4})\over \Gamma((2q_i+5)/4)}[i{(B^{(i)})^2\over 8}]^{2q_i+1\over 4} \right\} \, \left(1+{\cal{O}}((B^{(i)})^2)\right).
\end{eqnarray}

\noindent One can check that the first term in the right hand side is identical to $I_{q_i-{1\over 2}}(0)$ as it should.  This result is interesting in view of the crude approximations we had been relying on in Ref.~\cite{QCD1} and \cite{QCD-II}.  In particular, the term of $\exp-(b/x)$ of Eq.~(\ref{Eq:28}) was expanded at leading linear order in $b \sim g\varphi(b) \ll 1$.  One can now see that, were it not for the terms at $q_i=0$, this expansion would yield the correct leading order estimate of Eq.~(\ref{Eq:12}).

One has now to deal with the remaining integrations, as they show up in Eqs.~(\ref{Eq:1}) and (\ref{Eq:26}), that is, firstly with the subsequent integrations of (dropping the primes of the $V'^i_{1,2}$)
\begin{eqnarray}\label{Eq:34}
\int_{-\infty}^{+\infty} {{{\rm{d}}V_1^i\over E_1}}\ e^{-i{\alpha_1^iV_1^i\over E-p}}\int_{-\infty}^{+\infty} {{{\rm{d}}V_2^i\over E_2}}\ e^{-i{\alpha_2^iV_2^i\over E+p}}\,G^{30}_{03}\left(-i{[g\varphi(b)V_2^iV_1^i]^2\over 128N_c\, E^2p^2} \biggr| {2q_i+1\over 4}, {1\over 2},0\right).
\end{eqnarray}

\noindent Eq.~(\ref{Eq:34}) is obtained by promoting the original $n$-component vectors $\alpha_1, \alpha_2$ to $N$-component ones,  defining ${\widehat{\alpha}}_i= (1,1,1,1)\otimes \alpha_i$, for $i=1,2$. Then, taking Eq.~(\ref{Eq:5}) into account, the first line of Eq.~(\ref{Eq:26}) can be written as
\begin{equation}
E_1^{-N}E_2^{-N}\int{\mathrm{d}^N {\hat{\alpha}}'_1}\, e^{i{\hat{\alpha}}'_1\cdot\, {\cal{OT}}} \int{\mathrm{d}^N {\hat{\alpha}}'_2}\, e^{i{\hat{\alpha}}'_2\cdot\, {\cal{OT}}}\int{\mathrm{d}^N V'_1}\, e^{-i{{{\hat{\alpha}}'_1\cdot\, V'_I}\over E-p}} \int{\mathrm{d}^N V'_{2} \, e^{-i{{{\hat{\alpha}}'_2\cdot\, V'_2}\over E+p}} },
\end{equation}

\noindent where ${\cal{O}}$ is the orthogonal matrix of Eq.~(\ref{O}); eventually, in Eq.~(\ref{Eq:34}), the {\textit{prime}}-superscripts are dropped, and likewise, the little hats.

 Rather than integrating a Meijer's function expansion that doesn't terminate, one can take advantage of {\textit{formula}} 20.5.(8) of Ref.~\cite{Erdelyi1954}, so as to obtain for Eq.~(\ref{Eq:34}) the exact result of
\begin{eqnarray}\label{Eq:35}
{m^2\over E^2}\,{4\pi \over \alpha_1^i\alpha_2^i} \, G^{32}_{43}\left( \left. {-i\over 8N_c}\left({g\varphi(b)\over  \alpha_1^i\alpha_2^i }\right)^2{m^4\over E^2p^2} \right|
\begin{array}{cccc}
  \frac{1}{2}, & \frac{1}{2}, & 0, & 0 \\
  \frac{2q_i+1}{4}, & \frac{1}{2}, & 0, &
\end{array}
\right)\,,
\end{eqnarray}
where the $CMS$-relation $E_1=E_2=E$ is used.

Then, in view of Eq.~(\ref{Eq:1}), two extra integrations remain to be carried out: Defining ${\cal{T}}$ the $N$-vector of matrices $(T, T,T,T)=(1,1,1,1)\otimes T$, where as in Eqs.~(\ref{Eq:1}) and (\ref{Eq:26}), $T$ stands for the $n$-vector of generators $T^a$, $a=1,\dots,n$ in the adjoint representation, one can write for (\ref{Eq:26}) without absolute values
\begin{eqnarray}\label{Eq:36}
& & {m^2\over E^2}\int_{-\infty}^{+\infty} {\rm{d}}\alpha_1^i\ e^{-i\alpha_1^i({\cal{OT}})_i/2}\int_{-\infty}^{+\infty} {\rm{d}}\alpha_2^i\ e^{-i\alpha_2^i({\cal{OT}})_i/2}\nonumber \\  & & \quad \times\,{4\pi\over \alpha_1^i\alpha_2^i} \, G^{32}_{43}\left( \left. {-i\over 8N_c}\left({g\varphi(b)\over  \alpha_1^i\alpha_2^i }\right)^2{m^4\over E^2p^2} \right|
\begin{array}{cccc}
  \frac{1}{2}, & \frac{1}{2}, & 0, & 0 \\
  \frac{2q_i+1}{4}, & \frac{1}{2}, & 0 &
\end{array}
\right)\,,
\end{eqnarray}

\noindent where ${\cal{O}}$ is the orthogonal matrix introduced in Eq.~(\ref{O}). For the sake of further integrations, Eq.~(\ref{Eq:36}) can be most conveniently re-written as
\begin{eqnarray}\label{Eq:37}
&&{m^2\over E^2}\int_{-\infty}^{+\infty} {\rm{d}}\alpha_1^i\ e^{-i\alpha_1^i({\cal{OT}})_i/2}\int_{-\infty}^{+\infty} {\rm{d}}\alpha_2^i\ e^{-i\alpha_2^i({\cal{OT}})_i/2} \nonumber\\ & & \quad \times\,{4\pi\over \alpha_1^i\alpha_2^i} \,  G^{23}_{34}\left( \left. {8N_c\over -i}\left({ \alpha_1^i\alpha_2^i \over g\varphi(b) }\right)^2{E^2p^2\over m^4} \right|
\begin{array}{cccc}
  \frac{3 - 2 q_i}{4}, & \frac{1}{2}, & 1, &  \\
   \frac{1}{2}, & \frac{1}{2}, & 1, & 1
\end{array}
\right)\,,
\end{eqnarray}

\noindent where use has been made of the inversion formula {{5.3.1}}(9) of Ref.~\cite{Erdelyi1953},
\begin{eqnarray}\label{Eq:38}
G^{mn}_{pq} \left( x^{-1} \biggr|
\begin{array}{c}
  a_r(j) \\
  b_s
\end{array}
\right)
= G^{nm}_{qp}\left( x \biggr|
\begin{array}{c}
  1 - b_s \\
  1 - a_r(j)
\end{array}
\right).
\end{eqnarray}

\noindent By symmetry, Eq.~(\ref{Eq:37}) reads indeed as
\begin{eqnarray}\label{Eq:39}
&& -16\pi{m^2\over E^2}\int_0^{+\infty} {\rm{d}}\alpha_1^i\ {\sin[\alpha_1^i({\cal{OT}})_i]\over \alpha_1^i}\int_0^{+\infty} {\rm{d}}\alpha_2^i\ {\sin[\alpha_2^i({\cal{OT}})_i]\over \alpha_2^i} \nonumber\\  & & \quad \times \, G^{23}_{34}\left( \left.{8N_c\over -i}\left({ \alpha_1^i\alpha_2^i \over g\varphi(b) }\right)^2{E^2p^2\over m^4} \right|
\begin{array}{cccc}
  \frac{3 - 2 q_i}{4}, & \frac{1}{2}, & 1, &  \\
   \frac{1}{2}, & \frac{1}{2}, & 1, & 1
\end{array}
\right).
\end{eqnarray}

\noindent Considering the product of the $N$- integrations, and the Trace to be taken over internal color degrees, the first line of Eq.~(\ref{Eq:39}) leads to the expression
\begin{equation}\label{p}
(-16\pi{m^2\over E^2})^N\ C\!\int{\rm{d}}p_1\dots\int{\rm{d}}p_{N(N-1)/2}\, f(p)\, \Tr\,\prod_{i=1}^N\,\sum_{k'_i,k_i=0}^\infty [\,({\cal{O}}^{ij}(p){\cal{T}}_j)^2\,]^{k'_i+k_i+1}\,,
\end{equation}
which results from the expansion of the 2 {\it{sine}}-functions of Eq.~(\ref{Eq:39}), and where the normalization constant in Eq.~(\ref{Eq:10}) reads
\begin{equation}
C^{-1}= \int{\rm{d}}p_1\  ..\ {\rm{d}}p_{N(N-1)/2}\ f(p_1,\dots, p_{N(N-1)/2})\,.
\end{equation}

\noindent By construction, the ${\cal{T}}^i$ are such that, $\forall i=1,\dots,N $, $\exists\, a=1,\dots, n$, such that ${{\cal{T}}^i}={T^a}$. Note that were it not for this average over the extra parameters $p_l$ introduced in Eq.~(\ref{Eq:10}), which does not factor out because of the $p_l$-dependence of the orthogonal matrices ${\cal{O}}(p)$, then, instead of Eq.~(\ref{p}), one got an identically null result under the form of
\begin{equation}\label{nul}
\ \Tr\,\prod_{i=1}^N\, ({\cal{T}}_j)^2=\Tr\,\left(\prod_{a=1}^n\, (T^a)^2\right)^4=\Tr\,\prod_{a=1}^n\, (T^a)^2=0\,,
\end{equation}

\noindent where these equalities come from the fact that in the $N$-plet of matrices ${\cal{T}}$, each of the $n$- $T^a$ appears $D=4$ times by construction, from the idempotency of the $(T^a)^2$, and to the fact that they commute, $[(T^a)^2,(T^b)^2]=0$. A complete, exact result for Eq.~(\ref{p}) is not straight forward and will be presented elsewhere~\cite{TG2013}.  It can be checked that Eq.~(\ref{p}) comes out non-vanishing and, in agreement with Lattice accurate calculations~\cite{Lattice}, not exactly proportional to the first Casimir operator eigenvalue $C_2(R)$, over any given gauge group representation, $R$. For the time being, we will here content ourselves with the leading contribution to Eq.~(\ref{p}), so that Eqs.~(\ref{Eq:1}) and (\ref{Eq:26}) (without absolute values) read eventually
\begin{eqnarray}\label{Eq:40}
&& C^{st}\,(-16\pi {m^2\over E^2})^N\,C_F\,\prod_{i=1}^N \int_0^{+\infty} \left(\prod_{J=1}^2{\rm{d}}\alpha_J^i\ {\sin({{}}\alpha_J^i)\over \alpha_J^i}\right)\nonumber \\  & & \quad \times \, G^{23}_{34}\left( \left. {8N_c\over -i}\left({ \alpha_1^i\alpha_2^i \over g\varphi(b) }\right)^2{E^2p^2\over m^4} \right|
\begin{array}{cccc}
  \frac{3 - 2 q_i}{4}, & \frac{1}{2}, & 1, &  \\
   \frac{1}{2}, & \frac{1}{2}, & 1, & 1
\end{array}
\right),
\end{eqnarray}

\noindent where $C_F$ is the standard quadratic Casimir operator eigenvalue on the gauge group fundamental representation. Recalling the shorthand $Dn=N$, and getting back to Eq.~(\ref{56}), now integrated over the $V_{1,2}^i$, one obtains therefore a $4$-point fermionic amplitude proportional to
\begin{eqnarray}\label{75}
&& C^{st} (-{m^2\over E^2})^N\,C_F\left({64\pi^2\over g^2}\right)^{N\over 4}\nonumber \\ \nonumber & & \quad \times \sum_{\mathrm{monomials}}(\pm 1)\prod_{\sum q_j=N(N-1)/2}^{1\leq j\leq N}\ [1-i(-1)^{q_j}]\int_{0}^{+\infty} {\rm{d}}\alpha_1^j\ {\sin{{}}\alpha_1^j\over \alpha_1^j} \nonumber\\  & & \quad \times \int_{0}^{+\infty} {\rm{d}}\alpha_2^j\ {\sin{{}}\alpha_2^j\over \alpha_2^j} \, G^{23}_{34}\left( \left. {8N_c\over -i}\left({ \alpha_1^j\alpha_2^j \over g\varphi(b) }\right)^2{E^2p^2\over m^4} \right|
\begin{array}{cccc}
  \frac{3 - 2 q_j}{4}, & \frac{1}{2}, & 1, &  \\
   \frac{1}{2}, & \frac{1}{2}, & 1, & 1
\end{array}
\right).
\end{eqnarray}

\noindent In the second line, the sum runs over the monomials composing the Vandermonde determinant, and the $(\pm 1)$ is here to mean that each monomial appears with a plus or minus sign.
Already at $N=4$, one gets $64$ monomials of overall degree $6$ in the the eigenvalues $\xi_i$. But at $N=6$ one gets $32768$ monomials of overall degree $15$ that require several pages to be written down. One need not specify the number of monomials entailed in the Vandermonde determinant at $N=12$, relevant to the case $N_c=2$, not to speak of $N_c=3$, for which $N=32$.
\par\medskip
Before proceeding to the full integration of Eq.~(\ref{75}), it is worth observing that the last line,
\begin{eqnarray}\label{Eq:42}
\int_0^\infty {{\rm{d}}x\over x}\sin{x} \, \cdot G^{23}_{34}(C^{st}{x^2\over g^2\varphi^2}|\dots)
\end{eqnarray}

\noindent was the essential part of our first and crudely approximated estimate in Ref.~\cite{QCD1}, and the result was
\begin{eqnarray}\label{Eq:43}
I(g\varphi)={4\over g\varphi}\int_0^{g\varphi} {{\rm{d}}x\over x}\sin( x).
\end{eqnarray}

\noindent Based on Eq.~(\ref{Eq:43}), we had found for $SU(2)$ that at large values of $g\varphi(b)$, corresponding to small impact parameters, $I(g\varphi)\sim 1/g\varphi$, obviously damped in comparison to the large impact parameter limit of $g\varphi \ll 1$, where $I(g\varphi)\sim C^{st}$. The interpretation proposed in Ref.~\cite{QCD1} and based on Eq.~(\ref{Eq:43}), preserves its essential features within the present paper's more accurate calculation where one gets a behavior of $1/g\varphi$ at large impact parameter, and a behavior of ${\sqrt{g\varphi}}$ at small impact parameter.

Focusing on a given integration, that of index $j$ for example, one gets for the integrals of Eq.~(\ref{75}),
\begin{eqnarray}\label{Eq:44}
\int_{0}^{\infty} {\rm{d}}x\ {\frac{\sin{\left[x\sqrt{{\frac{g\varphi{\sqrt{-i}}}{\sqrt{8N_c}}} \frac{m^2}{Ep}}\right]}}{x}} \int_{0}^{\infty} {\rm{d}}y\ \frac{\sin{\left[y\sqrt{{\frac{g\varphi{\sqrt{-i}}}{\sqrt{8N_c}}} \frac{m^2}{Ep}}\right]}}{y} \, G^{23}_{34}\left(  x^2 y ^2 \biggr|
\begin{array}{c}
  a_r(j) \\
  b_s
\end{array}
\right)
\end{eqnarray}

\noindent with $\{ .. a_r(j)\ ..\}$ and $\{ .. b_s\ ..\}$ the sequences of numbers such as displayed in Eq.~(\ref{75}). These two sequences of numbers are such that the reduction formula {{5.3.1.}}(7) of Ref.~\cite{Erdelyi1953}, can be used to simplify somewhat the Meijer's special function entering Eq.~(\ref{Eq:44}) and write
\begin{eqnarray}\label{Eq:45}
G^{23}_{34}\left(  x^2 y^2 \biggr|
\begin{array}{cccc}
  \frac{3 - 2 q_j}{4}, & \frac{1}{2}, & 1 &  \\
  \frac{1}{2}, & \frac{1}{2}, & 1, & 1
\end{array}
\right)=G^{22}_{23}\left( x^2 y^2 \biggr|
\begin{array}{ccc}
  \frac{1}{2}, & \frac{3 - 2 q_j}{4} &    \\
  \frac{1}{2}, & \frac{1}{2},        & 1
\end{array}
\right).
\end{eqnarray}

\noindent This allows one to re-write Eq.~(\ref{Eq:44}) as
\begin{eqnarray}\label{Eq:46}
\int_{0}^{\infty} {\rm{d}}x\ {\sin[x\sqrt{{{g\varphi{\sqrt{-i}}\over{\sqrt{8N_c}}}}{m^2\over Ep}}]}\int_{0}^{\infty} {\rm{d}}y\ {\sin[y\sqrt{{{g\varphi\sqrt{-i}\over{\sqrt{8N_c}}}}{m^2\over Ep}}]}\,G^{22}_{23} \left( x^2 y^2 \biggr\vert
\begin{array}{ccc}
  0, & \frac{1 - 2 q_j}{4} &    \\
  0, & 0,                  & \frac{1}{2}
\end{array}
\right),
\end{eqnarray}

\noindent where the following identity, {{5.3.1.}}(8) of Ref.~\cite{Erdelyi1953}, has been used also,
\begin{eqnarray}\label{Eq:47}
x^\sigma G^{mn}_{pq}\left( x \biggr\vert
\begin{array}{c}
  a_r(j) \\
  b_s
\end{array}
\right)=G^{mn}_{pq}\left( x \biggr\vert
\begin{array}{c}
  a_r(j) + \sigma \\
  b_s + \sigma
\end{array}
\right).
\end{eqnarray}

In the end, parameters are such, in the {\textit{sine}} function as well as in the Meijer special function, the sequences of numbers $\{ \cdots \, a_r(j)+\sigma \, \cdots \}$ and $\{ \cdots \, b_s+\sigma \, \cdots \}$, that ${{20.5.}}(7)$~\cite{Erdelyi1954}, can be used twice so as to yield for Eq.~(\ref{Eq:46}) the result
\begin{eqnarray}\label{Eq:48}
{\pi{\sqrt{8N_c}}\over g\varphi(b)\sqrt{-i}}\,{Ep\over m^2} \ G^{42}_{36}\left( \left({  g\varphi(b)\sqrt{-i}\over {\sqrt{128N_c}}  }{m^2\over Ep}\right)^2 \biggr|
\begin{array}{cccccc}
  1, & 1, &             & \frac{1}{2}                        &              &  \\
  1, & 1, & 1,          & \frac{2 q_j +3}{4}, & \frac{1}{2}, & \frac{1}{2}
\end{array}
\right),
\end{eqnarray}

\noindent where the inversion formula {{5.3.1}}(9) of Ref.~\cite{Erdelyi1953}, that is, Eq.~(\ref{Eq:38}), has been used again in order to express the result in a form suited to a $ g\varphi(b) \ll 1$ limit expansion of the net amplitude.

Gathering all factors, the amplitude under consideration, Eq.~(\ref{Eq:26}) without absolute values, can eventually be displayed under the form of
\begin{eqnarray}\label{normalEq:49}
&& \nonumber  C_F\left(-{\sqrt{1-{4m^2\over {\widehat{s}}}}}\right)^N\times \sum_{\mathrm{monomials}}(\pm 1)\prod_{\sum q_j=N(N-1)/2}^{1\leq j\leq N}\ [1-i(-1)^{q_j}]  \\  && \left({{\sqrt{8iN_c}}\over g\varphi(b)}\right)\times\  G^{42}_{36}\left(  \left({  g\varphi(b)\over {\sqrt{8iN_c}}  }{m^2\over{\sqrt{{\widehat{s}}({\widehat{s}}-4m^2)}}}\right)^{2} \biggr| 
\begin{array}{cccccc}
  1, & 1, &             & \frac{1}{2}                        &              &  \\
  1, & 1, & 1,          & \frac{2 q_j +3}{4}, & \frac{1}{2}, & \frac{1}{2}
\end{array}
\right),
\end{eqnarray}

\noindent which relies on the normalization of Eq.~(\ref{Eq:11'}) (in particular, this normalization has the effect of eliminating the instanton-like contributions of ${64\pi^2/ g^2}$ that appear in (\ref{75})). A notation typical of perturbative Drell-Yan scattering processes has been introduced, ${\widehat{s}}=(p_1+p_2)^2$, the quark's $4$-momenta, the $p_i$, being related to their respective hadronic carriers in the usual {\textit{partonic}} way, $p_i=x_i{{P}}_i$, with $x_i\in[0,1]$. Beyond the notation itself, this is certainly far from an innocuous fact, as an explicit link shows up between a perturbative {partonic} content, essentially controlled by $\hat{s}$ and $m^2$, and a hadronic non-perturbative component controlled by $g\varphi(b)$, Eq.~(\ref{newphi}), and this, within one and the same overall amplitude expression: This is surely an important realization of the effective locality analysis~\cite{deTeranmond2012}.

It is worth noticing that the small $g\varphi(b)$-limit expansion of the $G^{42}_{36}$-Meijer's functions is the more relevant as their argument is not $g\varphi(b)$ itself, but rather the combination $C^{st}\,g\varphi(b)\times  m^2/{\widehat{s}}$, that is a small number in the eikonal and light quark mass limit. At leading order, corresponding to the value $1/2$ of the $b_s$- Meijer's function parameters, the two last terms of Eq.~(\ref{normalEq:49}) combine into a constant, $g\varphi(b)$-independent contribution, exactly as it should (see Eq.~(\ref{limit}) below).


Worth observing also is the interplay between the $2$-body scattering {\textit{probing energy}} $\hat{s}$, and the non-perturbative physics controlled by the function $g\varphi(b)$ (it is to be recalled that the large coupling limit is considered here, $g \gg 1$).  The Meijer's special functions argument, in effect, reads basically as
\begin{equation}\label{argument}
C^{st}\times g\times {\mu\over{\sqrt{\hat{s}}}}\,e^{-(\mu b)^{2+\xi}}\times {m^2/ {\widehat{s}}\over{\sqrt{1-{4m^2/ {\widehat{s}}}}}},
\end{equation}

\noindent so that the relative magnitude of $\hat{s}$ with respect to the scale $\mu$, decides of either the aperture (at ${\mu/{\sqrt{\hat{s}}}}\geq 1$), or of the closure (at ${\mu/{\sqrt{\hat{s}}}} \ll 1$), of the non-perturbative physics channel that is related to the $\mu$-dependences of Eq.~(\ref{normalEq:49}). In the latter case in effect, the Meijer's special functions behaviour is such that one has
\begin{eqnarray}\label{limit}
& & {{\sqrt{8iN_c}}\over g\varphi(b)}\times G^{42}_{36}\left([  {  g\varphi(b)\over {\sqrt{8iN_c}}  }{m^2/ {\widehat{s}}\over{\sqrt{1-{4m^2/ {\widehat{s}}}}}}]^{2} \biggr| 
\begin{array}{cccccc}
  1, & 1, &             & \frac{1}{2}                        &              &  \\
  1, & 1, & 1,          & \frac{2 q_j +3}{4}, & \frac{1}{2}, & \frac{1}{2}
\end{array}
\right) \nonumber \\ &=& {m^2/ {\widehat{s}}\over {\sqrt{1-{4m^2/ {\widehat{s}}}}}}+{\cal{O}}(\sqrt{g\varphi}),
\end{eqnarray}

\noindent whose leading order term is free of interaction dependences, exactly as it should, while leaving a residual non-perturbative and sub-leading effect on the order of $(g\mu/\sqrt{\hat{s}})^{1/2}\times(m^2/\hat{s})^{3/2}$, which, at high enough probing energy $\hat{s}$, can be quite small. Clearly, as is well known from Nuclear Physics, non-perturbative aspects are totally washed out beyond a certain energy scale (of $400$Mev~\cite{Kos}).

Also, it is interesting to observe that because the argument Eq.~(\ref{argument}) can become small in various ways, one has an indication that QCD should manifest {\textit{Asymptotic Freedom}} beyond the sole criterion of Perturbation Theory, as it has been advocated recently on the basis of Dyson-Schwinger Equations analyses~\cite{Trento}.

Eventually, whenever Eq.~(\ref{argument}) is not that small valued a combination, then an amplitude like Eq.~(\ref{normalEq:49}) describes a non-trivial blending of non-perturbative aspects together with perturbative degrees of freedom. These remarks easily extend to the cases of $2m$-point fermionic Green's functions, describing the $m$-bodies scattering processes of nuclei collisions, for example. In this situation, the overall energy of $s=(P_1+P_2)^2$ gets distributed among all of the possible $C_m^2$- $2$-body scattering sub-processes, and a summation over the sub-energy distributions complying with the constraint of global energy conservation must be performed.

That is, the concise form obtained for the fermionic correlation functions can certainly be exploited in various situations, limits, and along various numerical and/or semi-analytical ways. Similarly, when the full flavor and mass and structure function richness of QCD is restored, these generic forms of amplitudes could be used and tested at a phenomenological level to describe hadronic collisions at {\it{intermediate}} energies, for example. This however, will require that the absolute value prescriptions of Eq.~(\ref{Eq:26}) be circumvented in a way or other. In the present article one is interested in understanding quark's binding potential such as derived in Ref.~\cite{QCD-II}, based on the relationship between an eikonal amplitude and the potential able to generate it. As advertised in the main text of the present article, that result, given in Eq.~(\ref{Eq:S2-12}), displays the occurrence of an intriguing $\xi$-parameter whose non-zero value is essential for obtaining a non-trivial confining potential. Introducing a convenient representation for the Heavyside step function~\cite{TG2013}, we have been able to check that the absolute value prescriptions bring no change to the following analysis, and this is why they have been dropped in the current section.

As the derivation of Eq.~(\ref{Eq:S2-12}) relied upon frequently crude approximations, a natural issue is to know wether that $\xi$-parameter is an {\textit{avatar}} of the approximate calculational scheme used in Ref.~\cite{QCD-II}, or if it is not. It is this question that can be addressed now that, with respect to our first crude estimations~\cite{QCD1,QCD-II}, the calculational accuracy has been substantially improved.

Following the standard procedure that relates a 2-body eikonal amplitude to a potential acting between them~\cite{QCD-II,Matveev2002,HMF2}, it turns out that from Eq.~(\ref{normalEq:49}), the next-to-leading order of the $ g\varphi(b) \ll 1$ limit expansion is required.

Now, as made obvious by Meijer's functions small $ g\varphi(b)$ expansions, Eqs.~(\ref{Eq:31}) and (\ref{Eq:33}), the next-to-leading contributions are obtained from the parameters ${2q_j+3/ 4}$ whenever $q_j=0$, and instead of order $ g\varphi(b)$, as derived previously~\cite{QCD-II}, are indeed on the order of ${\sqrt{g\varphi(b)}}$.

However, as argued in Ref.~\cite{QCD-II}, it is the logarithm of that next-to-leading order contribution which is associated to the calculation of $V(r)$. In view of Eq.~(\ref{Eq:S2-10}) or (\ref{newphi}) though, one has $ \varphi(b)\sim e^{-(\mu b)^{2+\xi}}$ (see also Ref.~\cite{Matveev2002}, where a form of $\varphi(b)\sim e^{-b^2/4a}$ is proposed). Therefore, not solely that power of $1/2$, instead of $1$, but any other possible next-to-leading power of $g\varphi(b)$ would never bring the slightest change to the conclusion of Ref.~\cite{QCD-II}: That is, a non-trivial quark confining potential requires a non-vanishing $\xi$-parameter value.

In eikonal and quenched approximations at least, this conclusion is now reached on a substantially improved basis, and it is made accessible thanks to the non-perturbative effective locality property of QCD.

\section{\label{SEC:5}Discussion}

If effective locality is a genuine property of the non-perturbative fermionc sector of QCD, then it is worth exploring its consequences, that is, it is worth looking at QCD from the point of view of effective locality. It is the task that has been undertaken in the present article, where we have tried to make more rigorous the arguments of two previous analyses~\cite{QCD1,QCD-II}, while putting forth new interesting aspects.
\par\smallskip
In the first place, evaluating QCD fermionic amplitudes, it is argued that an awkward term of $\delta^{(2)}({\vec{b}})$, appearing as an overall multiplicative factor in an exponential argument, can not be avoided and is in no way related to any approximation {\textit{artifact}}. Inspection rather suggests that the occurrence of this factor should be understood as a compelling indication that the ordinary notion of an elementary particle cannot be maintained in the non-perturbative realm of QCD.


This term of $\delta^{(2)}({\vec{b}})$ was accordingly corrected, and, in a first attempt, turned into the smoother form of a normalized Gaussian distribution of transverse distance of the scattering quarks, \textit{i.e.}, their collision's impact parameter.  We could note that in an analysis of "Relativistic Quark Models in the Quasi-potential Approach", a similar form had been proposed~\cite{Matveev2002}.  Against all odds, though, the quark confining potential associated to the scattering eikonal amplitude is then found to be identically zero~\cite{QCD-II}.

Subsequently, an elementary \textit{deformation} of the random transverse quark gaussian distribution has been proposed in order to prevent such an odd result from occurring~\cite{QCD-II}. Fortunately enough, at $\xi<0$, that deformation (which, more precisely, doesn't bear on the L\'{e}vy flight distribution itself, but on its \textit{characteristic function}), falls into a stable and canonical generalization of random variable Gaussian distributions that are compatible with a generalized \textit{Central Limit Theorem} of Statistical Physics~\cite{StableDistribution}.

In a second place, we have made use of the most efficient {\textit{Random Matrix}} calculus so as to obtain the generic structure of fermionic QCD amplitudes, at least at quenched and eikonal approximations.  Illustrated on the simpler case of a $4$-point Green's function, the results can be easily extended to any $2n$-point function, and could therefore open the route to phenomenological applications.  An amazing and non-trivial aspect of the resulting amplitude is that within one and the same expression, it is able to encode some hadronic non-perturbative aspect together with the perturbative partonic content of the hadronic matter.  It is worth noting that such a link has been advocated recently (and the way by which this link must come about) within the AdS/QCD approach to non-perturbative QCD~\cite{deTeranmond2012}.

Moreover, the amplitude's generic forms so obtained comply with a conjecture concerning the form Quantum Field Theory's generating functionals must display in full generality~\cite{Ferrante2011}.  In effect, while the latter must be expressed in terms of {\textit{Fox special functions}}, our amplitudes turn out to be written in terms of {\textit{Meijer's special functions}} that are but a particular instance of the former.

However, the conclusions are here obtained on the bases of eikonal and quenched approximations to the full strong coupling fermionic scattering amplitudes. For example, this means that so long as quenched (at least) is not relaxed, the relevance of that $\xi$-parameter (describing a L\'evy-flight {\textit{Markovian}} diffusion mode of propagation for confined quarks) remains somewhat uncertain; and this, in spite of its fascinating relation to confinement, to a possible transverse non-commutative geometrical aspect of non-perturbative QCD, to confined quarks L\'evy-flight modes of propagation, to (spontaneous) chiral symmetry breaking and another related aspect of non-perturbative QCD not evoked in this article.

Since, on the other hand, the results of Ref.~\cite{Ferrante2011} are not restricted to any sort of approximation, one may hope that, at least at a qualitative level, our present conclusions could be somewhat preserved while relaxing the quenched approximation.  Still, that remains to be checked.  It is true also that the traceless character of the random matrices $M$ has not been implemented in our analysis: So far as could be seen, this point doesn't seem to modify in a drastic way the general features put forth in this article; still, it needs to be evaluated in a cogent way.  Clearly, more work is necessary to explore on firm grounds the consequences of such an intriguing, non-perturbative property as effective locality.

In these perspectives, quark propagator's full calculation and self energy effects should bring some new understanding, and shall be the matter of forthcoming analyses.

%
%
%

\appendix
\section{Proof of Eq.~(\ref{Eq:30})}

Starting from $I_p(b)$ such as given in Eq.~(\ref{Eq:28}) and evaluated thanks to Ref.~\cite{Apelblat1983}, at $p>0$ and $b>0$,
\begin{equation}\label{Eq:A1}
I_p(b):=\int_0^\infty\ {dx}\ x^p\ e^{-x^2-{b\over x}}={1\over 2{\sqrt{\pi}}} G^{30}_{03}\left({b^2\over 4}| {p+1\over 2}, {1\over 2},0\right),
\end{equation}

\noindent one proceeds to calculate Eq.~(\ref{Eq:27}), that is the integral
\begin{equation}\label{Eq:A2}
\int^{+\infty}_{-\infty}\ {dx\over {\sqrt{x}}}\ x^{q_i}\ e^{-{i}x^2-{ A^{(i)}\over x}}.
\end{equation}

\noindent This is first
\begin{equation}\label{Eq:A3}
\int_0^\infty\ {dx}\ x^{q_i-{1\over 2}}\ e^{-{i}x^2-{A^{(i)}\over x}}-i(-1)^{q_i}\int_0^\infty\ {dx}\ x^{q_i-{1\over 2}}\ e^{-{i}x^2-{-A^{(i)}\over x}},
\end{equation}

\noindent and can be re-written as
\begin{equation}\label{Eq:A4}
\left(\int_0^\infty\ {dx}\ x^{q_i-{1\over 2}}\ e^{-ix^2-{A^{(i)}\over x}}-i(-1)^{q_i}\int_0^\infty\ {dx}\ x^{q_i-{1\over 2}}\ e^{-ix^2-{-A^{(i)}\over x}}\right).
\end{equation}

\begin{figure}
\ifpdf
\includegraphics[scale=0.5]{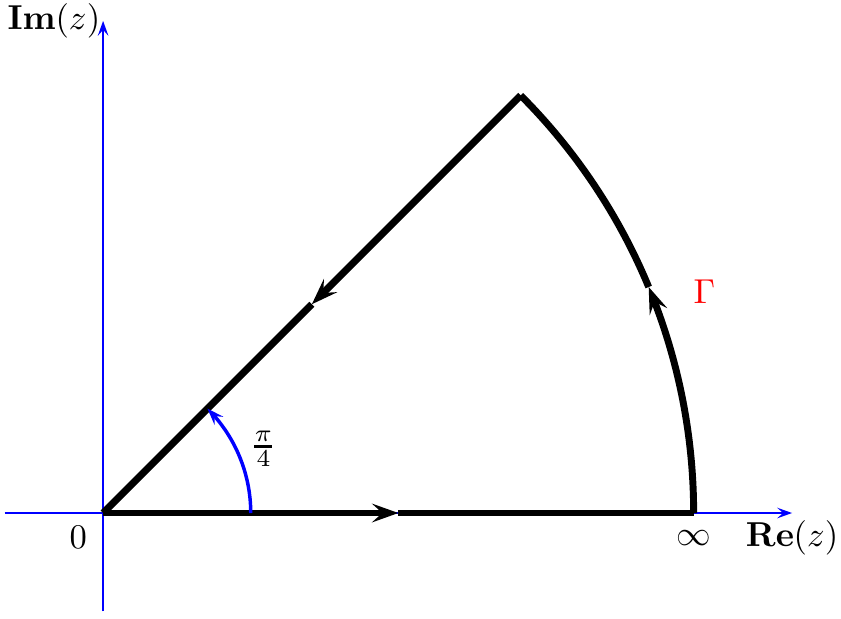}%
%
  %
%
\else
 \begin{pspicture}(10,10)
  \psline[linecolor=blue]{->}(1,2)(9,2)
  \psline[linecolor=blue]{->}(2,1)(2,7)
  \psline[linewidth=2pt]{->}(2,2)(5,2)
  \psline[linewidth=2pt]{-}(5,2)(8,2)
  \psarc[linewidth=2pt]{->}(2,2){6}{0}{22.5}
  \psarc[linewidth=2pt]{-}(2,2){6}{22.5}{45}
  \psline[linewidth=2pt]{->}(6.242641,6.242641)(4.12132,4.12132)
  \psline[linewidth=2pt]{-}(4.12132,4.12132)(2,2)
  \psarc[linecolor=blue,linewidth=1pt]{->}(2,2){1.5}{0}{45}
  \rput[l](3.5,2.65){$\frac{\pi}{4}$}
  \rput(1.75,1.75){$0$}
  \rput(9,1.75){$\mathbf{Re}(z)$}
  \rput(1.5,7){$\mathbf{Im}(z)$}
  \rput(8,1.75){$\infty$}
  \rput[l](8.0,4.25){\color{red} $\Gamma$}
 \end{pspicture}
\fi
\caption{\label{Fig:1Caption}Integration contour $\Gamma$.}
\end{figure}
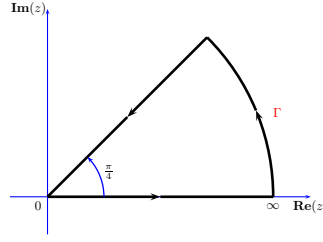

\noindent Now the first integral of Eq.~(\ref{Eq:A4}) can be evaluated by means of Cauchy's theorem:
\begin{equation}\label{Eq:A5}
\int_\Gamma\ {dz}\ z^p\ e^{-z^2-{c\over z}},
\end{equation}

\noindent where the integration contour $\Gamma$ is as depicted in Fig.~\ref{Fig:1Caption}, and where, to begin with, $p$ must be restricted to positive rationals. In this way, zero is obtained as
\begin{equation}\label{Eq:A5b}
I_p(c) - {\sqrt{i}}^{p+1}\int_0^\infty\ {dx}\ x^p\ e^{-ix^2-{c/{\sqrt{i}}\over x}}=0.
\end{equation}

\noindent Now, the analyticity of Meijer's special functions in their arguments, as well as the possibility of continuing their parameters to arbitrary complex values~\cite{Erdelyi1953} allows one to define the required integral as
\begin{equation}\label{Eq:A6}
\int_0^\infty\ {dx}\ x^{q_i-{1\over 2}}\ e^{-ix^2-{A^{(i)}/{\sqrt{8N_c}}\over x}}={{\sqrt{-i}}^{q_i+{1\over 2}}\over 2{\sqrt{\pi}}}G^{30}_{03}\left({i(A^{(i)})^2\over 4}| {2q_i+1\over 4}, {1\over 2},0\right),
\end{equation}

\noindent and Eq.~(\ref{Eq:A2}) eventually under the form of
\begin{equation}\label{Eq:A7}
\int^{+\infty}_{-\infty}\ {dx\over {\sqrt{x}}}\ x^{q_i}\ e^{-{i}x^2-{ A^{(i)}\over x}}={\sqrt{-i}}^{q_i+{1\over 2}}{[1-i(-1)^{q_i}]\over 2{\sqrt{\pi}}}G^{30}_{03}\left({i(A^{(i)})^2\over 4}| {2q_i+1\over 4}, {1\over 2},0\right),
\end{equation}

\noindent that is Eq.~(\ref{Eq:30}) in the main text.

%
%
%



%
%
%




%
%
%

\end{document}